\documentclass[fleqn,twoside]{article}
\usepackage[headings]{espcrc2}

\usepackage{graphicx}

\def\slash#1{\setbox0=\hbox{$#1$}\dimen0=\wd0                    
\setbox1=\hbox{/} \dimen1=\wd1 \ifdim\dimen0>\dimen1
\rlap{\hbox to \dimen0{\hfil/\hfil}} #1  \else                                       
\rlap{\hbox to \dimen1{\hfil$#1$\hfil}}  
      /   \fi}                                         

\newcommand{\be}{\begin{equation}}
\newcommand{\ee}{\end{equation}}
\newcommand{\bea}{\begin{eqnarray}}
\newcommand{\eea}{\end{eqnarray}}
\newcommand{\nn}{\nonumber}

\newcommand{\DB}{\Delta B}

\newcommand{\mev}{\,{\rm MeV}}   
   
\newcommand{\bdbd}{B_d-\overline B_d}
\newcommand{\bsbs}{B_s-\overline B_s}
\newcommand{\bqbq}{B_q-\overline B_q}

\title{B-Meson Mixing and Lifetimes}

\author{C.~Tarantino\address[roma3]{Dipartimento di Fisica, Universit\`a di Roma
Tre, and INFN, Sezione di Roma III, Via della Vasca Navale 84, I-00146 Rome, 
Italy }}
       
\begin{document}

\begin{abstract}
$B_d$ and $B_s$ meson systems play a fundamental role to test and improve our
understanding of Standard Model flavor dynamics.
The mixing parameters $\Delta m_{d,s}$ represent important constraints in the
unitarity triangle analysis.
Their theoretical estimates require non-perturbative calculations of B-meson 
decay constants and B-parameters; accurate results, recently obtained from the 
lattice, are reviewed.
Other phenomenologically interesting quantities are the beauty hadron lifetime
ratios and, the width differences and CP-violation parameters in $B_d$ and $B_s$
systems.
We discuss their theoretical predictions which, in the last four years, have
been improved thanks to accurate lattice calculations and next-to-leading order
perturbative computations.
\end{abstract}

\maketitle

\section{Introduction}
\label{intro}
The neutral $B_d$ and $B_s$ mesons mix with their antiparticles leading to 
oscillations between the mass eigenstates. The time evolution of the neutral 
meson doublet is described by a Schr\"odinger equation with an effective
$2 \times 2$ Hamiltonian
\bea
i\frac{d}{dt} \pmatrix{B_q\cr {\overline B}_q}=\qquad\quad\nn\\
\left[ \pmatrix{M_{11}^q & {M_{21}^q}^*\cr M_{21}^{q} & M_{11}^q}
-\frac{i}{2} \pmatrix{\Gamma_{11}^q & {\Gamma_{21}^q}^*\cr \Gamma_{21}^{q}& 
\Gamma_{11}^q} \right]\pmatrix{B_q\cr {\overline B}_q}.
\hspace{-1.cm}
\label{eq:schro}
\eea
The mass and width differences are defined as 
$\Delta m_q=m^q_H-m^q_L$ and $\Delta\Gamma_q=\Gamma^q_L-\Gamma^q_H$,
where $H$ and $L$ denote the Hamiltonian eigenstates with the heaviest and
lightest mass eigenvalue. 
These states can be written as
\be
\vert B_q^{L,H}\rangle ={1\over \sqrt{1+\vert (q/p)_q \vert^2}}\,\left(
\vert B_q\rangle \pm  \left(q/p\right)_q\vert {\overline B}_q\rangle
\right).
\ee

Theoretically, the hadron lifetime is related to $\Gamma^q_{11}$ 
($\tau_{B_q}=1/\Gamma^q_{11}$), while the observables $\Delta m_q$,
$\Delta \Gamma_q$ and $\vert(q/p)_q\vert$ are related to $M^q_{21}$ and 
$\Gamma^q_{21}$.
In $B_{d,s}$ systems, the ratio $\Gamma^q_{21}/M^q_{21}$ is of
${\cal O}(m_b^2/m_t^2)\simeq 10^{-3}$. Therefore, by neglecting terms of ${\cal 
O}(m_b^4/m_t^4)$, one can write
\bea
&\Delta m_q= 2\,\vert M^q_{21}\vert\,,\quad  
\Delta \Gamma_q=-2\,\vert M^q_{21}\vert\,
{\mathrm{Re}}\left(\frac{\Gamma^q_{21}}{M^{q}_{21}}\right)\,,&\nn\\
&\left\vert\left(q/p\right)_q\right\vert= 1+\frac{1}{2}\,{\mathrm{Im}}\left(
\frac{\Gamma^q_{21}}{M^q_{21}}\right)\,.&
\label{eq:dgammared}
\eea

The matrix elements $M^q_{21}$ and $\Gamma^q_{21}$ are related, respectively, 
to the dispersive and the absorptive parts of the $\DB=2$ transitions. 
In the SM, these transitions are the result of second-order charged weak 
interactions involving the well-known box diagrams.

The dispersive matrix element $M^q_{21}$ has been computed at 
the NLO in QCD~\cite{Buras:1990fn}.
Since the mass differences $\Delta m_q$ play a fundamental role in constraining 
the unitarity triangle, it is important to have precise theoretical predictions
of $M^q_{21}$.
Recently,  accurate studies of the B-meson decay constants and B-parameters
entering in $M^q_{21}$, have been performed on the lattice. 
 
The absorptive matrix elements $\Gamma^q_{11}$  and 
$\Gamma^q_{21}$ can be computed by applying the heavy quark expansion 
(HQE)~\cite{ope}, with a consequent separation of the
short-distance contributions from the long-distance ones.
The great energy ($\sim m_b$) released in beauty hadron decays, in fact, allows 
to expand the inclusive widths in powers of $1/m_b$.
Theoretical predictions of inclusive rates are based on a 
non-perturbative calculation of matrix elements, widely studied in lattice QCD,
 and a perturbative calculation of Wilson coefficients.

Recently, the contribution of light quarks in beauty hadron decay widths 
(spectator effect) has been computed at $\mathcal{O}(\alpha_s)$ in QCD and 
$\mathcal{O}(\Lambda_{QCD}/m_b)$ in the HQE.
Based on these calculations is the theoretical prediction for beauty 
hadron lifetimes and B-meson width differences and CP-violation parameters. 
Improved theoretical estimates have been
obtained, to be compared with recent accurate experimental measurements or
limits.


\section{Mixing parameters}
\label{sec:0}
The mass difference in the $\bdbd$ system ($\Delta m_d$) is proportional to
$f^2_{B_d} \hat B_{B_d} |V_{td}|^2$, thus representing a constraint on the CKM
element $|V_{td}|$, provided that the multiplied hadronic matrix elements are
calculated.
The analogous mass difference in the $\bsbs$ system ($\Delta m_s$) can be used
to reduce the theoretical uncertainty, by considering the ratio
\be
\frac{\Delta m_s}{\Delta m_d} \propto  \frac{|V_{ts}|^2}{|V_{td}|^2} \xi^2 \quad
{\mathrm with}\quad\xi = \frac{f_{B_s} \sqrt{\hat B_{B_s}}}{f_{B_d} 
\sqrt{\hat B_{B_d}}}\,.
\label{eq:uta0}
\ee

The hadronic parameter that is better determined from lattice calculations is 
$f^2_{B_s} \hat B_{B_s}$, whereas $\xi$ and  $f^2_{B_d} \hat
B_{B_d}$ are affected by larger uncertainties coming from the chiral
extrapolations.
These uncertainties are strongly correlated.
For this reason, the best approach to constraint the unitarity triangle, 
recently proposed and adopted in Ref.~\cite{UTA}, uses the following equations 
\bea
\qquad\Delta m_d \propto \frac{f^2_{B_s} \hat B_{B_s}}{\xi^2}\,,\qquad\qquad
\frac{\Delta m_s}{\Delta m_d} \propto \xi^2 \,.
\label{eq:uta1}
\eea

In lattice calculations the difficulty to treat heavy quarks has been
essentially solved by introducing the HQET based lattice actions, and the
results from different formulations are in good agreement, in the quenched
approximation, within the systematic uncertainty of 
$15$\%~\cite{Hashimoto:2004hn}.
Concerning $f_{B_s}$, unquenched results seem slightly higher than the 
quenched values ($\sim 10-15$\%)~\cite{Hashimoto:2004hn} but some systematics, 
as continuum scaling, are not yet investigated.
In the case of the $\hat B_{B_q}$ parameter, instead, sea quark effects result
to be negligible.
Concerning the chiral extrapolation, it represents a delicate issue for
$f_{B_d}$ and, nowadays, chiral log effects are finally estimated 
($\sim 5$\%)~\cite{Hashimoto:2004hn}, while chiral loops are not a problem for 
$\hat B_{B_d}$, as expected from ChPT.

In Table~\ref{tab:summary} we summarize the averages for the $\bqbq$ mixing
parameters, which include rough estimates of chiral logs and unquenched effects.
This year, new results have been obtained by the HPQCD 
Collaboration~\cite{hpqcd} that confirm their previous 
estimates~\cite{hpqcd_old}.
The averages in Table~\ref{tab:summary}, therefore, are identical to those 
presented at ICHEP2004 by S.~Hashimoto~\cite{Hashimoto:2004hn}.

\begin{table}
  \centering
  \begin{tabular}{cc}
    \hline
& Averages\\
    \hline
    $f_{B_d}$ &  $189(27) \mev$ \\
    $f_{B_s}$ & $230(30) \mev$ \\
    $f_{B_d}\hat{B}_{B_d}^{1/2}$ & $214(38) \mev$ \\
    $f_{B_s}\hat{B}_{B_s}^{1/2}$ & $262(35) \mev$ \\
    $f_{B_s}/f_{B_d}$ & $1.22$($^{+5}_{-6}$) \\
    $\xi$ & $1.23(6)$ \\
    \hline
  \end{tabular}
  \label{tab:summary}
\vspace*{-0.5cm}
\end{table}


\section{Beauty hadron lifetime ratios}
\label{sec:1}

The experimental values of the measured lifetime ratios of beauty hadrons 
are~\cite{BBpage}
\bea
\label{eq:rexp}
&\frac{\tau(B^+)}{\tau(B_d)}=1.081 \pm 0.015 \,, 
\frac{\tau(B_s)}{\tau(B_d)}=0.939 \pm 0.044 \,,& \nn\\
&\frac{\tau(\Lambda_b)}{\tau(B_d)}=0.803 \pm 0.047\,.&
\eea

By applying the HQE, the inclusive decay width of a hadron $H_b$ can be
expressed as  a sum of local $\DB=0$ operators of increasing dimension, as
\be
\qquad\Gamma (H_b) = \sum_k \frac{\vec{c}_k (\mu)}{m_b^k}\,\langle H_b | 
\vec{O}^{\DB = 0}_k (\mu) | H_b \rangle\,.
\label{eq:sum}
\ee
The HQE yields the separation of short distance effects, confined in the Wilson 
coefficients ($\vec{c}_k$), from long distance physics, represented by the 
matrix elements of the local operators ($\vec{O}_k^{\DB=0}$).

Spectator contributions, which distinguish different beauty 
hadrons, appear at $\mathcal{O}(1/m_b^3)$  in the HQE.
These effects, although suppressed by an additional power of $1/m_b$, are 
enhanced with respect to leading contributions by a phase-space factor of 
$16 \pi^2$, being $2 \to 2$ processes instead of $1 \to 3$ 
decays~\cite{Bigi:1992su,NS}.
In order to evaluate the spectator effects one has to calculate the matrix 
elements of dimension-six current-current and penguin operators, 
non-perturbatively, and their Wilson coefficients, perturbatively.

Concerning the perturbative part, the NLO QCD corrections to the coefficient
functions of the current-current operators have been
computed~\cite{Ciuchini:2001vx}-\cite{NOI}.

Concerning the non-perturbative part, the non-valence contributions,
corresponding to contractions of two light quarks in the same point,
have not been computed. Their non-perturbative lattice calculation would be 
possible, in principle, however it requires to deal with the difficult problem 
of power-divergence subtractions.
On the other hand, the valence contributions, which exist when the light quark
of the operator enters as a valence quark in the external hadronic state, have 
been evaluated. 
For $B-$mesons, QCD and HQET lattice results have been recently combined
to extrapolate to the physical $b$ quark mass~\cite{APE}, while for the 
$\Lambda_b$ baryon, lattice-HQET has been used~\cite{DiPierro:1999tb}.
These accurate results are in agreement with the values obtained in previous
lattice studies~\cite{Becirevic:2001fy}-\cite{DiPierro:1998cj}.

Last year, the sub-leading spectator effects which appear at 
$\mathcal{O}(1/m_b^4)$ in the HQE, have been included in the analysis of 
lifetime ratios.
The relevant operator matrix elements have been estimated in the vacuum 
saturation approximation (VSA) for $B-$mesons and in the quark-diquark model 
for the $\Lambda_b$ baryon, while the corresponding Wilson coefficients have 
been calculated at the leading order (LO) in QCD~\cite{Gabbiani:2003pq}.

\begin{figure}
\includegraphics[width=18pc]{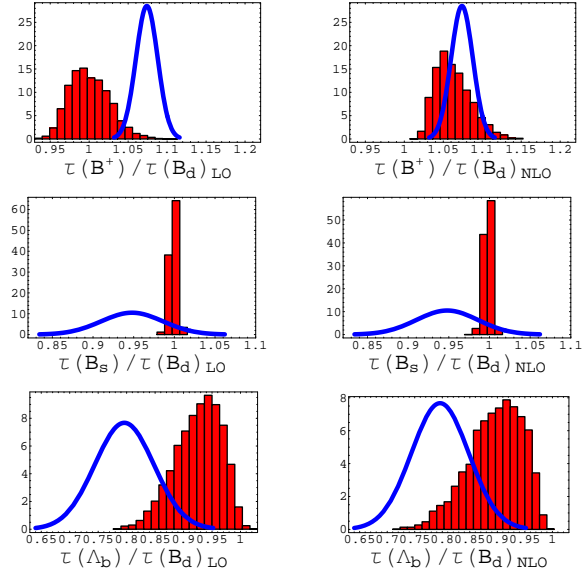}
\vspace*{-0.9cm}
\caption{\it Theoretical (histogram) vs experimental (solid line) 
distributions of lifetime ratios. The theoretical predictions are shown 
at the LO (left) and NLO (right).}
\label{fig:plot}
\end{figure}

Updated theoretical predictions for the lifetime ratios are~\cite{NOI}
\bea
\label{eq:nlores}
&\left. \frac{\tau(B^+)}{\tau(B_d)} \right|_{\rm NLO}\!\!\!\!\!\!\!\!\!\! =
1.06 \pm 0.02 \,, 
\left. \frac{\tau(B_s)}{\tau(B_d)} \right|_{\rm NLO}\!\!\!\!\!\!\!\!\!\! =
1.00 \pm 0.01 \,,&  \nn\\
&\left. \frac{\tau(\Lambda_b)}{\tau(B_d)} \right|_{\rm NLO}\!\!\!\!\!\!\!\!\!\! =
0.88 \pm 0.05\,.&
\eea
They turn out to be in good agreement with the experimental measurements of 
Eq.~(\ref{eq:rexp}).

It is worth noting that the agreement at $1.2 \sigma$ between the theoretical
prediction for the ratio $\tau(\Lambda_b)/\tau(B_d)$ and its experimental value
is achieved thanks to the inclusion of NLO (see Fig.~\ref{fig:plot}) and the
$1/m_b$ corrections to spectator effects. 
They both decrease the central value of $\tau(\Lambda_b)/\tau(B_d)$ by $8$\% and
$2$\% respectively.

Further improvement of the $\tau(\Lambda_b)/\tau(B_d)$ theoretical prediction 
would require the calculation of the current-current operator non-valence 
B-parameters and of the perturbative and non-perturbative contribution of the 
penguin operator, which appears at the NLO and whose matrix elements present the
same problem of power-divergence subtraction.
These contributions are missing also in the theoretical predictions of 
$\tau(B^+)/\tau(B_d)$ and $\tau(B_s)/\tau(B_d)$, but in these cases they 
represent an effect of $SU(2)$ and $SU(3)$ breaking respectively, and are 
expected to be small.

\section{Neutral $B_q$-meson width differences}
\label{sec:2}
The width difference between the ``light''  and ``heavy'' neutral $B_q$-meson 
($q=d, s$) is defined in terms of the off-diagonal matrix element 
$\Gamma^q_{21}$ (see Eq.~(\ref{eq:dgammared})).

In the HQE of $\Gamma^q_{21}$, the leading contribution comes at 
$\mathcal{O}(1/m_b^3)$ and is given by dimension-six $\DB=2$ operators.
Up to and including $\mathcal{O}(1/m_b^4)$ contribution, one can write 
\bea
\Gamma^q_{21} =
-\frac{G_F^2 m_b^2}{24 \pi M_{B_q}}\cdot
\left[
c^q_1(\mu_2) {\langle \overline B_q \vert {\cal O}^q_1(\mu_2) \vert B_q\rangle}
+ \right.\nn\\
\left.c^q_2(\mu_2) {\langle \overline B_q \vert {\cal O}^q_2(\mu_2) \vert B_q\rangle} 
+ \delta^q_{1/m}\right]\, ,
\label{eq:gamma12q}
\eea
where ${\langle \overline B_q \vert {\cal O}^q_i(\mu_2) \vert B_q\rangle}$ are the 
matrix elements of the two independent dimension-six operators, $c^q_i(\mu_2)$
their Wilson coefficients, known at the NLO in 
QCD~\cite{BBlargh}-\cite{Beneke:2003az}, while $\hat{\delta}_{1 / m_b}$ 
represents the contribution of the dimension-seven operators~\cite{seven}.

Lattice results of the dimension-six operator matrix 
elements~\cite{Gimenez:2000jj}-\cite{Lellouch:2000tw} have been confirmed and 
improved, by combining QCD and HQET results in the heavy quark 
extrapolation~\cite{damir}.
Moreover, the effect of the inclusion of the dynamical quarks has been examined,
within the NRQCD approach, finding that these matrix elements are essentially 
insensitive to switching from $n_f=0$ to 
$n_f=2$~\cite{Yamada:2001xp,Aoki:2003xb}.

Concerning the dimension-seven operators, their matrix elements have never been 
estimated out of the VSA.
Two of these four matrix elements, however, can be related through Fierz 
identities to the complete set of operators studied in Ref.~\cite{damir}.

\begin{figure}
\includegraphics[width=18pc]{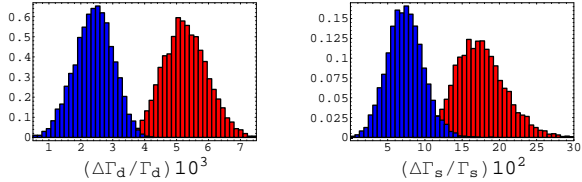}
\vspace*{-0.9cm}
\caption{\it Theoretical distributions for $B_d$ and
$B_s$ width differences, at the LO (light/red) and NLO (dark/blue).}
\label{fig:plot1}
\end{figure}

The updated theoretical predictions are~\cite{NOIwip}
\bea
\qquad\qquad\Delta \Gamma_{d}/\Gamma_{d} &=& (2.42 \pm 0.59)10^{-3} ,\nn\\
\Delta \Gamma_{s}/\Gamma_{s} &=& (7.4 \pm 2.4)10^{-2} \,.
\label{eq:dg_noi}
\eea
The corresponding theoretical distributions are shown in Fig.~\ref{fig:plot1}, 
where the effect of NLO corrections can be seen to be quite relevant.

The interest in $\Delta \Gamma_{s}$ has largely increased due to recent
experimental measurements from the CDF~\cite{Acosta:2004gt} and D0~\cite{Abazov:2005sa} 
Collaborations.
Previous experimental limits~\cite{BBpage}
\bea
\Delta \Gamma_{d}/\Gamma_{d} &=& 0.008 \pm 0.037 \pm 0.019 \,,\quad
\nn\\
\Delta \Gamma_{s}/\Gamma_{s} &=& 0.07^{+0.09}_{-0.07}\,,
\eea
were in agreement with theoretical predictions, within the large experimental 
uncertainties.
Last year CDF and D0 presented their results for $\Delta
\Gamma_{s}/\Gamma_s$~\cite{Acosta:2004gt,Abazov:2005sa}
\bea
\Delta \Gamma_{s}/\Gamma_{s} &=& 0.65^{+0.25}_{-0.33} \pm 0.01 (\mathrm{CDF})\,,
\nn\\
\Delta \Gamma_{s}/\Gamma_{s} &=& 0.21^{+0.27}_{-0.40} (\mathrm{D0})\,.
\eea
The CDF's result is surprisingly large, $2 \sigma$ away from the theoretical
prediction.
Updated experimental results with higher statistics are therefore needed for
a significant comparison.
The theoretical value, instead, is under control and comes from cancellations
occurring at the NLO and $\mathcal{O}(1/m_b^4)$, which successively reduce the 
LO central value from $0.26$ to $0.18$ and to the final $0.074$ given in 
Eq.~(\ref{eq:dg_noi}).

Our theoretical prediction is slightly smaller than the value calculated in
Ref.~\cite{BBlargh} ($\Delta \Gamma_{s}/\Gamma_{s}=0.12(5)$).
 This difference is mainly due to the contribution of  $\mathcal{O}(1/m_b^4)$
 which, in Ref.~\cite{BBlargh} is wholly estimated in the VSA, while we express
the matrix elements of two dimension-seven operators in terms of those 
calculated on the lattice.

\section{CP Violation parameters: $\vert(q/p)_d\vert$ and $\vert(q/p)_s\vert$}
\label{sec:3}

The experimental observable $\vert(q/p)_q\vert$, whose deviation from unity
describes CP-violation due to mixing, is related to $M_{21}^q$ and 
$\Gamma_{21}^q$, through Eq.~(\ref{eq:dgammared}).
The theoretical prediction of $\vert(q/p)_q\vert$ is therefore based on the same
perturbative and non-perturbative calculation discussed in Sec.~\ref{sec:2}, 
while the $V_{CKM}$ contribution to $\vert(q/p)_q\vert$ is different from that 
in  $\Delta \Gamma_q /\Gamma_q$.

\begin{figure}
\includegraphics[width=18pc]{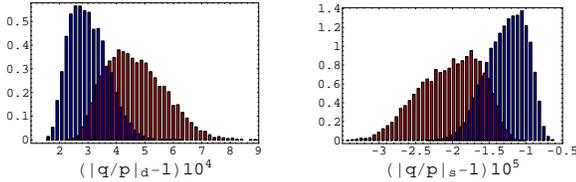}
\vspace*{-0.9cm}
\caption{\it Theoretical distributions for $\vert(q/p)_q\vert-1$ in $B_d$ and  
$B_s$ systems, at the LO (light/red) and NLO (dark/blue).}
\label{fig:plot3}
\end{figure}

The updated theoretical predictions~\cite{NOIwip} are
\bea
\label{eq:nlores3}
\vert(q/p)_d\vert-1&=&(2.96 \pm 0.67) 10^{-4}\,,\nn\\
\vert(q/p)_s\vert-1&=&-(1.28 \pm 0.28) 10^{-5}\,.
\eea
The corresponding theoretical distributions are shown in Fig.~\ref{fig:plot3},
with an evident effect of NLO corrections.

A preliminary measurement for $\vert(q/p)_d\vert -1$  is now available from the 
BABAR Collaboration~\cite{BBpage}
\be  
\vert (q/p)_d\vert -1=0.029\pm 0.013\pm 0.011
\ee
Improved measurements are certainly needed to make this comparison 
significant for the unitarity triangle analysis.

It is a pleasure to thank D.~Becirevic, M.~Ciuchini, E.~Franco, V.~Lubicz and
F.~Mescia for sharing their insights in topics covered by this talk.
I would also like to thank the Beauty2005 organizers for the pleasant
conference realized in Assisi. 


\end{document}